\documentclass[submission]{eptcs}
\providecommand{\event}{INFINITY 2011} 

\usepackage{amsmath}
\usepackage{amssymb}
\usepackage{amsthm}
\usepackage{ascmac}
\usepackage{latexsym}
\usepackage[dvipdfm]{graphicx}

\newcommand{\tuple}[1]{\langle #1 \rangle}

\newtheorem{thm}{Theorem}

\newtheorem{prp}[thm]{Proposition}

\newtheorem{dfn}[thm]{Definition}

\theoremstyle{remark}
\newtheorem{note}{Note}

\title{A Probabilistic Temporal Logic with Frequency Operators\\ and Its Model Checking}
\author{
Takashi Tomita \qquad \qquad Shigeki Hagihara \qquad \qquad Naoki Yonezaki\\
\institute{Dept.\ of Computer Science,}
\institute{Graduate School of Information Science and Engineering,}
\institute{Tokyo Institute of Technology}
\email{\{tomita, hagihara, yonezaki\}@fmx.cs.titech.ac.jp}
}

\def\titlerunning{A Probabilistic Temporal Logic with Frequency Operators and Its Model Checking}
\def\authorrunning{T. Tomita, S. Hagihara \& N. Yonezaki}

\begin{document}
\maketitle


\begin{abstract}
Probabilistic Computation Tree Logic (PCTL) and
 Continuous Stochastic Logic (CSL)
 are often used to describe specifications of probabilistic properties for discrete time and continuous time, respectively.
In PCTL and CSL,
 the possibility of executions satisfying some temporal properties
 can be quantitatively represented by the probabilistic extension of the path quantifiers
 in their basic Computation Tree Logic (CTL),
 however,
 path formulae of  them are expressed via the same operators in CTL.
For this reason,
 both of them cannot represent formulae with quantitative temporal properties,
 such as those of  the form
 ``some properties hold to more than 80\% of time points (in a certain bounded interval) on the path.''
In this paper,
 we introduce
 a new temporal operator which expressed the notion of frequency of events,
 and
 define probabilistic frequency temporal logic (PFTL) based on CTL$^*$.
As a result,
 we can easily represent the temporal properties of behavior in probabilistic systems.
However,
 it is difficult to develop a model checker for the full PFTL, due to rich expressiveness.
Accordingly,
 we develop
 a model-checking  algorithm for the CTL-like fragment of PFTL against finite-state Markov chains,
 and
 an approximate model-checking algorithm for the bounded 
 Linear Temporal Logic (LTL) -like fragment of PFTL against countable-state Markov chains.
\end{abstract}

%
%
\section{Introduction}\label{intro}

To analyze probabilistic systems,
 probabilistic model checking is often used.
In probabilistic model checking,
 the inputs are a probabilistic model and a probabilistic property described in a specification language,
 and
 the output is whether or not the model satisfies the property.
Probabilistic Computation Tree Logic \cite{hansson1994,kwiatkowska2007} (PCTL) and
 Continuous Stochastic Logic \cite{aziz1996, baier2003,kwiatkowska2007} (CSL)
 are often used to describe specifications of probabilistic properties.
PCTL and CSL are probabilistic extensions of
 Computation Tree Logic 
 \cite{emerson1990} (CTL)
 for discrete-time and continuous-time, respectively.
In PCTL and CSL,
 the probabilistic path quantifier ${\bf P}$ is introduced in place of 
 the universal path quantifier ${\bf A}$ (for all paths, $\ldots$) and
 the existential path quantifier ${\bf E}$ (there exists a path such that $\ldots$).
As a result,
 we can quantitatively represent the possibility of executions satisfying temporal properties of interest.
However,
 PCTL and CSL can only describe path formulae with temporal operators
 of the form ``some properties hold in the next state''
 via the next-operator ${\bf X}$;
 of the form  ``some properties eventually hold''
 via the eventually-operator ${\bf F}$ (or $\Diamond$);
 of the form  ``some properties always hold''
 via the always-operator ${\bf G}$ (or $\Box$);
 and
 of the form  ``some properties hold at a certain time point and other properties hold until that point''
 via the until-operator ${\bf U}$.
Thus,
 ``property $\varphi$ holds to more than 80\% of time points (in the interval $[0,10]$) on the path''
 cannot be represented in PCTL or CSL.
To capture similar quantitative properties of an above example,
 CSL additionally has the steady-state operator ${\bf S}$ \cite{baier2003,kwiatkowska2007},
 and there are also
 extensions of PCTL and CSL with reward (or cost) structure \cite{kwiatkowska2007}.
Even though,
 the steady-state operator ${\bf S}$
 can only capture the expected steady-state probability of being states satisfying properties of interest,
 and
 PCTL/CSL with rewards
 can only express the properties of the expected value of cumulated reward associated with states or transitions.

To capture temporal properties of this kind,
 it is necessary to employ the integral of the duration of states, as in Duration Calculus \cite{zhou1991} (DC).
In DC,  the above property is explicitly described by $\int_0^{10} \varphi(t) \text{d}t  \geq 8$.
In this paper,
 we describe such properties using the concept of frequency
 and
 introduce probabilistic frequency temporal logic (PFTL) based on CTL$^*$ \cite{emerson1990},
 for discrete-time/continuous-time.
To this logic,
 we add the (conditional) frequency operator ${\bf Q}$.
Using  the frequency operator $\bf Q$,
 we describe
 the above path property by
 ${\bf Q}_{>0.8}^{\leq 10} \varphi$
 in PFTL.
PFTL has rich expressiveness,
 and hence it is difficult to develop a model checker for the full logic
 (see Section \ref{model_checking}).
However,
 we develop a numerical model-checking algorithm
 for the CTL-like fragment of PFTL against finite-state Markov chains (MCs),
 and
 a statistical model-checking algorithm
 for the bounded Linear Temporal Logic \cite{emerson1990} (LTL)
 -like fragment of PFTL against infinite-state MCs.
The outline of the numerical algorithm for the CTL-like fragment
 is similar to that of PCTL and CSL \cite{hansson1994,baier2003,kwiatkowska2007}.
We compute transient and steady-state probabilities and reachability via matrix operations.
The difference is
 that our technique requires the number of states satisfying the formulae of interest to be counted in terms of frequency.
On the other hand,
 the statistical algorithm is an approximate one, based on statistical inference,
 and hence
 there are errors (although the significance level can be set according to our needs).
However,
 we anticipate that it will provide useful information in many cases.
%
%
%
We estimate whether or not ``an input MC satisfies an input formula''
 using the sequential probability ratio test \cite{wald1945} (SPRT), as in \cite{younes2002} for CSL.
%
%

The remainder of this paper is organized as follows.
In Section \ref{stochastic_model},
 we give the definitions of discrete-time/continuous-time MCs,
 and describe their probabilistic behavior.
In Section \ref{ptl},
 we define the syntax and semantics of PFTL
 and discuss the expressiveness of PFTL.
In Section \ref{model_checking},
 we present the numerical model-checking algorithm
 for the CTL-like fragment of PFTL against finite-state MCs,
 and
 the statistical model-checking algorithm
 for the bounded LTL-like fragment of PFTL against infinite-state MCs.
Our conclusions are stated in Section \ref{last}.

%
%
\section{Markov chains}\label{stochastic_model}

In this section,
 we present the definitions of discrete-time/continuous-time MCs
 and describe their probabilistic behavior.
We fix a set $AP$ of atomic propositions that expresses the properties of interest.
\begin{dfn}
A (labeled) discrete-time Markov chain (DTMC) $\mathcal{D}$ is a tuple $(S,\bar{s}, P, L)$
such that:
 $S$ is a countable set of states;
 $\bar{s} \in S$ is an initial state;
 $P : S^2 \to [0,1]$ is a transition probability matrix satisfying the condition that
 $\sum_{s' \in S} P(s,s')=1$ and
 $\{ s' | P(s,s')>0 \}$ is finite for all $s$;
 $L:S \to 2^{AP}$ is a labeling function
 that assigns to each state the set of valid atomic propositions in the state.
\end{dfn}
$P(s,s')$ denotes the probability of a one-step transition from $s$ to $s'$.
An execution (or discrete-time path) of
 a DTMC $\mathcal{D}$ is represented by an infinite sequence of states
 $\omega= s_0 s_1 \ldots$, where $\forall i. P(s_i,s_{i+1})>0$
 and
 $\Omega_s^{\mathcal{D}}$
 is the set of all paths starting from state $s$ in $\mathcal{D}$.
For a path $\omega = s_0 s_1 \ldots$,
 we denote the $i$-th state $s_i$ by $\omega(i)$ and
 the $i$-th suffix $s_i s_{i+1} \ldots$ by $\omega^{i}$.
Let
 $C_{s_0}^{\mathcal{D}}(s_0 \ldots s_n)$ be 
 a cylinder set
 $\{\omega \in \Omega_{s_0}^{\mathcal{D}} |\forall i \leq n. \omega (i) = s_i \}$,
 and
 let
 $\Sigma_{\Omega_{s_0}^{\mathcal{D}}}$ be
 the smallest $\sigma$-algebra
 containing all the cylinder sets $C_{s_0}^{\mathcal{D}}(s_0, \ldots , s_n)$
 in $\Omega_{s_0}^{\mathcal{D}}$.
The probability measure $Pr_{s_0}^{\mathcal{D}}$ on
 the measurable space $(\Omega_{s_0}^{\mathcal{D}}, \Sigma_{\Omega_{s_0}^{\mathcal{D}}})$ is
 uniquely defined as follows:
\begin{eqnarray*}
Pr_{s_0}^{\mathcal{D}}(C_{s_0}^{\mathcal{D}}(s_0 \ldots s_n))=
\displaystyle\prod_{i=1}^{n} P(s_{i-1},s_i).
\end{eqnarray*}
\begin{dfn}
A (labeled) continuous-time Markov chain (CTMC) $\mathcal{C}$ is a tuple $(S,\bar{s}, Q, L)$
 such that:
 $S$ is a countable set of states;
 $\bar{s} \in S$ is an initial state;
 $Q : S^2 \to \mathbb{R}$ is an infinitesimal generator matrix satisfying the condition that
 $\sum_{s' \in S\setminus \{s\}} Q(s,s')=-Q(s,s)$,
 $Q(s,s') \geq 0$ if $s\neq s'$
 and
 $\{ s' | Q(s,s')>0 \}$ is finite for all $s$;
 $L:S \to 2^{AP}$ is a labeling function
 that assigns to each state the set of valid atomic propositions in the state.
\end{dfn}
$Q(s,s')$ is the rate of a one-step transition from $s$ to $s'$ if $s\neq s'$.
Otherwise,
 $-Q(s,s)$ is the exit rate from $s$
 and
 the spent time in $s$ is exponentially distributed with parameter $-Q(s,s)$.
An execution (or continuous-time path) of
 a CTMC $\mathcal{C}$ is represented by
 an infinite alternating sequence $\omega= s_0 t_0 s_1 t_1 \ldots$
 or a finite and non-empty sequence $\omega= s_0 t_0 \ldots s_n \infty$,
 where $s_i \in S$ and $t_i \in \mathbb{R}_{>0}$
 (this value represents the time spent in $s_i$) for all $i \geq 0$.
 $\Omega_s^{\mathcal{C}}$
 is the set of all paths starting from state $s$ of $\mathcal{C}$.
For a path
 $\omega=s_0 t_0 s_1 t_1 \ldots$ ($s_n \infty$),
 we denote
 the $i$-th state $s_i$ by $\omega(i)$,
 the $i$-th spent time $t_i$ by $time(\omega,i)$ and
 the suffix $s_i t_i' s_{i+1} t_{i+1} \ldots$ after time point $t$ by $\omega^t$,
 where $i=\min \{ i' | \sum_{j=0}^{i'} t_{j} > t \}$ and $t_i' = \sum_{j=0}^{i} t_{j} - t$.
A path $\omega$ is called an infinite time-length path if $\sum_{i=0}^{\infty} time(\omega,i)=\infty$
 (therefore, an infinite number of transitions do not occur in any bounded intervals of $\mathbb{R}_{\geq 0}$ on the path).
For an interval $I$ in $\mathbb{R}_{\geq 0}$,
 let
 $C_{s_0}^{\mathcal{C}}(s_0, I_0, s_1, I_1, \cdots, I_{n-1}, s_n)$
 be
 a continuous-time cylinder set
 $\{ \omega \in \Omega_{s_0}^{\mathcal{C}}(s_0) | \omega(i)=s_i \wedge time(\omega,i) \in I_i \}$,
 and
 let
 $\Sigma_{\Omega_{s_0}^{\mathcal{C}}}$ be
 the smallest $\sigma$-algebra
 that contains all cylinder sets $C_{s_0}^{\mathcal{C}}(s_0, I_0,  s_1, I_1,\cdots, I_{n-1}, s_n)$ in $\Omega_{s_0}^{\mathcal{C}}$.
The probability measure $Pr_{s_0}^{\mathcal{C}}$ on
 the measurable space $(\Omega_{s_0}^{\mathcal{C}}, \Sigma_{\Omega_{s_0}^{\mathcal{C}}})$ is
 uniquely defined as follows:
\begin{eqnarray*}
Pr_{s_0}^{\mathcal{C}}(C_{s_0}^{\mathcal{C}}(s_0, I_0,  s_1, I_1,\cdots, I_{n-1}, s_n))=
\displaystyle \prod_{i=0}^{n-1}
\frac{Q(s_{i},s_{i+1})}{-Q(s_{i},s_{i})} \cdot \int_{I_{i}} -Q(s_{i},s_{i}) \cdot e^{Q(s_{i},s_{i}) t}\text{d}t.
\end{eqnarray*}
We assume that
 CTMCs in this paper are not explosive, that is, almost all paths of them are infinite time-length.

In numerical computations for CTMCs,
 transient probabilities are tractable
 and
 the uniformization method is a standard technique for computing transient probabilities of CTMCs.
\begin{dfn}
For a CTMC $\mathcal{C} = (S, \bar{s}, Q, L)$ such that $\sup\{-Q(s,s) | s \in S\}$ is finite,
a uniformized DTMC $\text{unif}_{\lambda}(\mathcal{C})$ is
$(S,\bar{s}, {\bf I} + Q/\lambda, L)$,
where 
$\lambda$ is a uniformization rate greater than or equal to $\sup\{-Q(s,s) | s \in S\}$
and ${\bf I}$ is the unit matrix.
\end{dfn}

If each transition time in $\text{unif}_{\lambda}(\mathcal{C})$ is exponentially distributed with parameter $\lambda$,
 the behavior of $\text{unif}_{\lambda}(\mathcal{C})$ is equivalent to that of $\mathcal{C}$ in a sense.
For a uniformized DTMC $\text{unif}_{\lambda}(\mathcal{C}) = (S,\bar{s}, P, L)$,
 the transient probability matrix $\Pi_{k}^{\mathcal{C}}$
 ($\Pi_{k}^{\mathcal{C}}(s,s')$ is the probability of being in state $s'$, $k$ time-units after the current state $s$ in $\mathcal{C}$)
 is computed as follows:
\begin{eqnarray*}
\Pi_{k}^{\mathcal{C}} = \sum_{n=0}^{\infty} \rho(n ;\lambda k) \cdot P^n
\text{ where $\rho(n; \lambda k)$ is the Poisson distribution $e^{- \lambda  k}  (\lambda  k)^n/n!$.}
\end{eqnarray*}
In the numerical computation,
 this infinite sum can be truncated.
%
The truncation points
 can be determined by Fox-Glynn algorithm \cite{fox1988},
 which gives an $\mathcal{O}(\lambda k)$-size upper bound.
%
%
%
\section{Probabilistic frequency temporal logic}\label{ptl}
In this section,
 we define the syntax and semantics of PFTL for discrete-time/continuous-time.
We discuss the expressiveness of PFTL in Section \ref{expressiveness}.
%
%
\subsection{Syntax and semantics}\label{pfctl}
\begin{dfn}
Probabilistic Frequency Temporal Logic (PFTL) is defined as follows:
\begin{eqnarray*}
{\text {state formula }} \varphi &::=&
 \ a \ 
 | \ \neg \varphi \ 
 | \ \varphi_1 \wedge \varphi_2 \ 
 | \  {\bf P}_{\sim p} [\psi]\\
{\text {path formula }}  \psi &::=&
 \varphi \ 
 | \ \neg \psi \ 
 | \ \psi_1 \wedge \psi_2 \ 
 | \ {\bf X} \psi \ 
 | \  \psi_1 {\bf U}^{I} \psi_2 \ 
 | \ {\bf Q}_{\bowtie q}^{I} \tuple{\psi_1 | \psi_2} \ 
\end{eqnarray*}
where
 $a \in AP$,
 $\sim, \bowtie \in \{ <,>,\leq,\geq \}$,
 $p,q \in [0,1]$ and
 $I$ is an interval of $\mathbb{N}$ for discrete time
 (or of $\mathbb{R}_{\geq 0}$ for continuous time).
\end{dfn}

Intuitively speaking,
 ${\bf P}_{\sim p} [\psi]$ means
 that the occurrence probability of paths starting from the given state and satisfying $\psi$ 
 obeys the bound $\sim p$;
 ${\bf X} \psi $ means
 that the suffix after the next state on the path satisfies $\psi$;
 $\psi_1 {\bf U}^{I}  \psi_2$  means
 that $\psi_2$ holds at a certain time point in the interval $I$ on the path and
 $\psi_1$ holds until that point is reached;
 ${\bf Q}_{\bowtie q}^{I} \tuple{\psi_1 | \psi_2} $ means
 that the conditional frequency of time points satisfying $\psi_1$ under the condition $\psi_2$
 in the interval $I$ on the path
 obeys the bound $\bowtie q$.
We allow the following abbreviations:
\begin{eqnarray*}
\varphi_1 \vee \varphi_2 &\equiv& \neg (\neg\varphi_1 \wedge \neg\varphi_2)\\
{\tt true} &\equiv& \varphi \vee \neg\varphi\\
\varphi_1 \rightarrow \varphi_2 &\equiv& \neg\varphi_1 \vee \varphi_2\\
{\bf Q}_{\bowtie q}^{I} \psi &\equiv&
{\bf Q}_{\bowtie q}^{I} \tuple{\psi | {\tt true}}\\
{\bf F}^{I} \psi &\equiv&
{\tt true} {\bf U}^{I} \psi\\
{\bf G}^{I} \psi &\equiv&
\neg {\bf F}^{I} \neg \psi
\end{eqnarray*}
In the sequel,
 we often omit the time bound $I$ if $I = [0,\infty)$
 and
 denote a time bound $\{i | i \bowtie k\}$ by $\bowtie k$
 and a time bound $\{j | j \pm i \in I \}$ by $I \mp i$.

We now describe the semantics for DTMCs.
The frequency in a finite interval is simply defined as
 the ratio of the number of time points satisfying subformulae in the interval.
For an unbounded ${\bold Q}$ formula,
 we write a semantics (called {\it limit semantics}) in terms of
 the limit superior and limit inferior of the global frequency on the path.
In general,
 the occurrence frequency of states in a path may not converge.
However,
 if the MC is finite,
 we can regard it as simply the limit of the global frequency of the path,
 because of the convergence property of the limit distribution of a finite-state MC.

\begin{dfn}
Let the DTMC $\mathcal{D} = (S,\bar{s},P, L)$.
For
 a state $s \in S$,
 a discrete-time path $\omega$,
 a state formula $\varphi$ and
 a path formula $\psi$,
the satisfaction relation $\models$ is defined as follows:
\begin{eqnarray*}
\mathcal{D}, s \models a & \Leftrightarrow & a \in L(s)\\
\mathcal{D}, s \models \neg \varphi & \Leftrightarrow &
 \mathcal{D}, s \not\models \varphi\\
\mathcal{D}, s \models \varphi_1 \wedge \varphi_2 & \Leftrightarrow &
 \mathcal{D}, s \models \varphi_1 \text{ and } \mathcal{D}, s \models  \varphi_2\\
\mathcal{D}, s \models {\bf P}_{\sim p} [\psi] & \Leftrightarrow &
 {Pr_s^{\mathcal{D}}(\{ \omega \in \Omega_s^{\mathcal{D}} | \mathcal{D}, \omega \models \psi \})} \sim p\\
\mathcal{D}, \omega \models \varphi & \Leftrightarrow &
 \mathcal{D}, \omega(0) \models \varphi \\
\mathcal{D}, \omega \models \neg \psi & \Leftrightarrow &
 \mathcal{D}, \omega \not\models \psi \\
\mathcal{D}, \omega \models \psi_1 \wedge \psi_2 & \Leftrightarrow &
 \mathcal{D}, \omega \models \psi_1 \text{ and } \mathcal{D}, \omega \models \psi_2 \\
\mathcal{D}, \omega \models {\bf X} \psi & \Leftrightarrow &
 \mathcal{D}, \omega^1 \models \psi \\
\mathcal{D}, \omega \models \psi_1  {\bf U}^{I} \psi_2 & \Leftrightarrow &
 \exists i \in I.  ( \mathcal{D},\omega^i \models \psi_2  \text{ and } 
 \forall j<i. \mathcal{D}, \omega^j \models \psi_1)
\end{eqnarray*}
\begin{eqnarray*}
\mathcal{D}, \omega \models {\bf Q}_{\bowtie q}^{I} \tuple{\psi_1 | \psi_2} & \Leftrightarrow &
\begin{cases}
{\tt true} & \textup{if } \{ i \in I | \mathcal{D}, \omega^i \models \psi_2 \} =\emptyset,\\
\displaystyle\frac{|\{ i \in I | \mathcal{D}, \omega^i \models \psi_1 \wedge \psi_2 \} |}
{|\{ i \in I | \mathcal{D}, \omega^i \models \psi_2 \} |}
 \bowtie q 
 &\textup{if } \sup I  \in \mathbb{N},\\
\displaystyle\limsup_{k \to \infty} 
\frac{|\{ i \in \mathbb{N}_{\leq k} \cap I | \mathcal{D}, \omega^i \models \psi_1 \wedge \psi_2 \} |}
{|\{ i \in \mathbb{N}_{\leq k} \cap I | \mathcal{D}, \omega^i \models \psi_2 \} |}
\bowtie q
 &\textup{if } \sup I =\infty \textup{ and } \bowtie \in \{<,\leq \},\\
\displaystyle\liminf_{k \to \infty}
\frac{|\{ i \in \mathbb{N}_{\leq k} \cap I  | \mathcal{D}, \omega^i \models \psi_1 \wedge \psi_2 \} |}
{|\{ i \in \mathbb{N}_{\leq k} \cap I | \mathcal{D}, \omega^i \models \psi_2 \} |}
\bowtie q
 &\textup{otherwise.}
\end{cases}
\end{eqnarray*}
\end{dfn}
We define a semantics for CTMCs as follows.
\begin{dfn}\label{sem_for_ctmc}
Let the CTMC $\mathcal{C} = (S,\bar{s},Q, L)$.
For
 a state $s \in S$,
 a continuous-time path $\omega$ with infinite time-length,
 a state formula $\varphi$ and
 a path formula $\psi$,
 the satisfaction relation $\models$ is defined as follows:
\begin{eqnarray*}
\mathcal{C}, s \models a & \Leftrightarrow & a \in L(s)\\
\mathcal{C}, s \models \neg \varphi & \Leftrightarrow &
\mathcal{C},  s \not\models \varphi\\
\mathcal{C}, s \models \varphi_1 \wedge \varphi_2 & \Leftrightarrow &
 \mathcal{C}, s \models \varphi_1 \text{ and } \mathcal{C}, s \models \varphi_2\\
\mathcal{C}, s \models {\bf P}_{\sim p} [\psi] & \Leftrightarrow &
{Pr_s^{\mathcal{C}}( \{ \omega \in \Omega_s^{\mathcal{C}} | \mathcal{C}, \omega \models \psi \})} \sim p\\
\mathcal{C}, \omega \models \varphi & \Leftrightarrow &
\mathcal{C}, \omega(0) \models \varphi \\
\mathcal{C}, \omega \models \neg \psi & \Leftrightarrow &
\mathcal{C}, \omega \not\models \psi \\
\mathcal{C}, \omega \models \psi_1 \wedge \psi_2 & \Leftrightarrow &
\mathcal{C}, \omega \models \psi_1 \text{ and } \mathcal{C}, \omega \models \psi_2 \\
\mathcal{C}, \omega \models {\bf X} \psi & \Leftrightarrow &
time(\omega,0) \in \mathbb{R}_{>0} \text{ and }
\mathcal{C}, \omega^{time(\omega,0)} \models \psi \\
\mathcal{C}, \omega \models \psi_1 {\bf U}^{I} \psi_2 & \Leftrightarrow &
 \exists t \in I.  ( \mathcal{C},\omega^t \models \psi_2 \text{ and } 
\forall t' \in (0,t). \mathcal{C}, \omega^{t'} \models \psi_1)\\
\mathcal{C}, \omega \models {\bf Q}_{\bowtie q}^{I} \tuple{\psi_1 | \psi_2} & \Leftrightarrow &
\begin{cases}
{\tt true} & \textup{if } \{ t \in I | \mathcal{C}, \omega^t \models \psi_2 \} =\emptyset, \\
f^I_{\tuple{\psi_1|\psi_2}}(0)
 \bowtie q
 &\textup{if } \sup I \in \mathbb{R},\\
\displaystyle\limsup_{k \to \infty}
f^{I \cap [0,k]}_{\tuple{\psi_1|\psi_2}}(0)
 \bowtie q
 &\textup{if } \sup I = \infty \text{ and } \bowtie \in \{<,\leq \},\\
\displaystyle\liminf_{k \to \infty}
f^{I \cap [0,k]}_{\tuple{\psi_1|\psi_2}}(0) 
\bowtie q
 &\textup{otherwise.}
\end{cases}
\end{eqnarray*}
where $f^{I}_{\tuple{\psi_1|\psi_2}}(t)$ is  the frequency of time points satisfying $\psi_1$ under $\psi_2$
 in the interval $I+t$, for the Lebesgue measure $\mathcal{L}$, given by: 
\begin{eqnarray*}
f^{I}_{\tuple{\psi_1|\psi_2}}(t) =
\begin{cases}
\displaystyle\frac{|\{ t' \in I +t | \mathcal{C}, \omega^{t'} \models \psi_1 \wedge \psi_2 \} |}
{|\{ t' \in I +t | \mathcal{C}, \omega^{t'} \models \psi_2 \} |}
& \textup{if } \sup I \neq \infty \textup{ and }
 \{ t' \in I +t | \mathcal{C}, \omega^{t'} \models \psi_2 \} \neq \emptyset \textup{ and} \\
& \quad \mathcal{L}(\{ t' \in I +t | \mathcal{C}, \omega^{t'} \models \psi_2 \})=0,\\
\displaystyle\frac{\mathcal{L}(\{ t \in I +t | \mathcal{C}, \omega^{t'} \models \psi_1 \wedge \psi_2 \})}
{\mathcal{L}(\{ t' \in I +t | \mathcal{C}, \omega^{t'} \models \psi_2 \})}
& \textup{if } \sup I \neq \infty \textup{ and }
 \mathcal{L}(\{ t' \in I +t | \mathcal{C}, \omega^{t'} \models \psi_2 \})>0,\\
{\tt undefined}
 & \textup{otherwise.}
\end{cases}
\end{eqnarray*}
\end{dfn}
In continuous time,
 we must consider two cases:
 the number of time points satisfying subformulae in a finite interval
 is
 either
 only finite
 or
 continuously infinite.
For finite time points,
 the frequency is defined in a manner similar to the discrete time.
For continuously infinite time points,
 the frequency is defined as the ratio of the Lebesgue measure of the set of time points satisfying subformulae.
By the following proposition (the proof is omitted from this paper),
 the set of time points satisfying subformulae is Lebesgue measurable.
It is not necessary to consider the case in which
 there exists a countably infinite number of time points satisfying subformulae in a finite interval.
\begin{prp}
For
 a CTMC $\mathcal{C}$,
 a path formula $\psi$,
 a bound interval $I$
 and
 a continuous-time path $\omega$ with infinite time-length,
 the set $\{ t \in I | \mathcal{C}, \omega^{t} \models \psi \}$ of time points satisfying $\psi$ in $I$ for $\omega$
 can be expressed as a finite union of intervals.
\end{prp}
\begin{note}
For an unbounded formula ${\bold Q}^I$ ($\sup I = \infty$),
 we can define alternative semantics as follows:
\begin{eqnarray*}
 \mathcal{D}, \omega \models {\bf Q}_{\bowtie q}^I \tuple{ \psi_1 | \psi_2} 
& \Leftrightarrow &
\exists i \in \mathbb{N}. \forall j>i. 
\frac
{|\{ j' \in I \cap [0,j] | \mathcal{D}, \omega^{j'} \models \psi_1 \wedge \psi_2 \} |}
{|\{ j' \in I \cap [0,j] | \mathcal{D}, \omega^{j'} \models \psi_2 \} |}
\bowtie q\\
 \mathcal{C}, \omega \models {\bf Q}_{\bowtie q}^I \tuple{ \psi_1 | \psi_2} 
& \Leftrightarrow &
\exists t \in \mathbb{R}. \forall t'>t. 
\frac
{\mathcal{L}(\{ t'' \in I \cap [0,t'] | \mathcal{C}, \omega^{t''} \models \psi_1 \wedge \psi_2 \} )}
{\mathcal{L}(\{ t'' \in I \cap [0,t'] | \mathcal{C}, \omega^{t''} \models \psi_2 \} )}
\bowtie q
\end{eqnarray*}
The above semantics (called {\it stable semantics}) would be tractable for analysis using automata-based methods,
because it is captured by the co-B\"{u}chi condition.
However,
 in the present paper,
 we use the limit semantics because it facilitates numerical model-checking.
\end{note}
%
%
\subsection{Expressiveness}\label{expressiveness}
PFTL can flexibly express properties of paths via the frequency operator ${\bf Q}$.
We present some examples and note the expressiveness of PFTL.
\begin{itemize}
\item \mbox{${\bf Q}_{>0} \psi$}: the global frequency of time points satisfying $\psi$ on a path is greater than $0$.
\begin{itemize}
\item
 This formula
 is not equivalent to
 ${\bf G} {\bf F} \psi$ representing ``$\psi$ is satisfied infinitely often on the path,''
 because
 the global frequency on the path may converge to $0$ even if $\psi$ is satisfied infinitely often.
\end{itemize}
\item \mbox{${\bf Q}_{>0.8}^{[0,20]} x=10$}:
 more than 80\% of the time points in $[0,20]$ satisfy the proposition $x=10$.
\begin{itemize}
\item
For probabilistic systems,
 states are often associated with numerical values as in MCs with rewards.
This formula is different than both ${\bf G}^{[0,20]} x=10$ and ${\bf G}^{[0,20]} 8 \leq x \leq 12$.
To capture behavior of a probabilistic system,
 we can write flexible expressions in PFTL. 
\end{itemize}
\item \mbox{${\bf P}_{= 1} [{\bf Q}_{\bowtie q} \varphi]$}:
the global frequency of the time points satisfying $\varphi$ obeys the bound $\bowtie q$ for almost all paths.
\begin{itemize}
\item
This formula is equivalent to the CSL formula ${\bf S}_{\bowtie q} [\varphi]$ 
 if the given MC is irreducible (that is, it is possible to reach any state from any state).
Otherwise, 
 the ${\bf S}$ formula means that the expected value of the global frequency obeys the bound $\bowtie q$.
\end{itemize}
\item \mbox{${\bf Q}_{>0.9} \tuple{\psi | \varphi}$}:
more than 90\% of time points satisfying $\varphi$ on the path also satisfy $\psi$.
\begin{itemize}
\item 
%
If we assume probabilistic fairness,
 this formula is similar to a path formula ${\bf G} (\varphi \to {\bf P}_{>0.9}[\psi])$
 that means
 the probabilistic branching property ${\bf P}_{>0.9}[\psi]$ holds at all states satisfying $\varphi$ on the path.
Furthermore,
 a conditional frequency (in a sense, it can be interpreted as a conditional probability)
 between path formulae on a path
 can be expressed via the ${\bf Q}$ operator
 without path quantifications.
%
\end{itemize}
\item \mbox{$\neg {\bold Q}_{>0.1} \varphi \wedge \neg {\bf Q}_{<0.9} \varphi$}:
the frequency of time points satisfying $\varphi$ becomes less than $0.1$ and also greater than $0.9$ infinitely often.
\begin{itemize}
\item
Roughly speaking,
 this formula describes a situation in which
 intervals where $\varphi$ frequently holds and
 intervals where $\varphi$ frequently  does not hold appear alternately
 and become progressively longer in both the limit semantics and the stable semantics.
However,
 it is not a property of the languages defined by $\omega$-Kleene closure, e.g., $\omega$-regular and $\omega$-context free languages.
In the discrete-time stable semantics,
 for natural numbers $q_1$ and $q_2$ such that $0<q_1<q_2$,
 a single frequency formula ${\bf Q}_{> q_1/q_2} \tuple{\varphi_1|\varphi_2}$ is a property of $\omega$-context free.
The class of $\omega$-context free language is equivalent to
 the class of language accepted by $\omega$-pushdown automata \cite{cohen77},
 and
 we can construct an $\omega$-pushdown automaton
 which stores the value $n \cdot (q_2-q_1) - m \cdot q_1$ in the stack, where $n$ and $m$ are the numbers of visiting states satisfying
 $\varphi_1 \wedge \varphi_2$ and $\neg\varphi_1 \wedge \varphi_2$, respectively.
Then ${\bf Q}_{> q_1/q_2} \tuple{\varphi_1|\varphi_2}$ can be represented by the automaton with
 the co-B\"{u}chi condition ``the stored value $n \cdot (q_2-q_1)- m \cdot q_1$ is non-positive at finitely many time-points.''
\end{itemize}
\end{itemize}
%
%
\section{Model checking}\label{model_checking}
In this section,
 we introduce model-checking algorithms.
The inputs are a DTMC $\mathcal{D}=(S,\bar{s},P, L)$
 (or a CTMC $\mathcal{C}=(S,\bar{s},Q, L)$)
 and a formula $\varphi$.
The output is
 whether or not $\mathcal{D}, \bar{s} \models \varphi$ (or $\mathcal{C}, \bar{s} \models \varphi$).
Unfortunately,
 it is difficult to develop a model-checking algorithm for PFTL
 because of its high expressiveness of path formulae, which describes linear time properties.
In the model checking of linear time logic against a (non-) probabilistic system,
 an automata-based approach is generally used.
In this type of approach,
 a (non-) deterministic $\omega$-automaton equivalent to (the negation of) the input path formula $\psi$
 is first constructed.
Then
 the synchronized product system of the input system and the constructed $\omega$-automata
 is analyzed.
Because
 the synchronized product system captures the intersection of the behavior of the input system and that (out) of $\psi$,
 we can reduce the model checking to the reachability (or emptiness) problem.
However,
 the language class of the path formulae in PFTL and its equivalent automata class
 are open in both the limit semantics and the stable semantics.
The limit semantics does not primarily match existing automata, which do not have an concept of convergence.

The stable semantics also results in intractable problems.
For discrete time,
 the language class of the path formulae in PFTL
 is at least a superclass of $\omega$-regular,
 and
 includes
 $\omega$-context free and non- $\omega$-regular languages and
 also non- $\omega$-Kleene closure languages.
Hence,
 for model checking using an automata-based approach,
 we require a new type of automata to capture frequency.
Such automata
 must have stack-like features,
 because they must be able to recognize some $\omega$-context free languages.
For continuous time,
 the set of the path formulae in PFTL
 is a superset of Metric Temporal Logic \cite{koymans90} (MTL),
 which is a real-time extension of LTL, in an interval-based semantics.
Timed automata \cite{alur1994} are widely used as real-time automata,
 however,
 there exist MTL formulae (including bounded formulae \cite{bouyer08})
 for which there is no equivalent timed automata.
We conjecture
 that the required automata to satisfy some frequency conditions in continuous time is
 some kind of extended timed automata
 and
 that it is also impossible to construct such a timed automaton to capture a property represented by a path formula in PFTL.
It may be possible to obtain a synchronized product directly,
 or
 it may not be necessary to employ an automata-based approach,
 but
 there is currently no available method for model checking of an LTL-like fragment of PFTL.

Accordingly,
 we develop separate model-checking algorithms for two fragments of PFTL.
The first is a strict numerical model-checking procedure
 for the CTL-like fragment of PFTL against finite-state MCs
 (Section \ref{for_numerical_model}).
The second is a statistics-based approximation model-checking
 for the bounded LTL-like fragment of PFTL against infinite-state MCs (Section \ref{for_statistical_model}).
The model checking for the bounded LTL-like fragment of PFTL against infinite-state DTMCs
 can be reduced to the model checking for LTL against finite-states DTMCs.
Because,
 the number of reachable states from the initial state for bounded steps is finite
 and
 a bounded ${\bf Q}^I$ formula can be translated into a nested ${\bf X}$ formula.
However,
 the translated formula has $\mathcal{O}(\inf I + 2^{|I|})$-size
 and hence
 it is difficult to check exactly for the bounded LTL-like fragment of PFTL in the viewpoint of complexity.
In a statistics-based approach,
 we sample prefix sequences of paths of an input MC by probabilistic simulation
 and
 statistically determine whether or not an input model satisfies an input formula by using the sample.
Thus,
 we can apply statistical methods to model checking for an infinite-state MC,
 because
 it is easy to generate prefix sequences of paths of an MC even if the MC has infinitely many states.
For finite-state MCs,
 numerical techniques are often limited by the state explosion problem.
Statistical methods can also overcome also this issue.
%
%
%
\subsection{Model checking of the CTL-like fragment of PFTL}\label{for_numerical_model}
In this section,
 we introduce a model checking algorithm for the CTL-like fragment of PFTL:
\begin{eqnarray*}
{\text {state formula }} \varphi &::=&
 a \ 
 | \ \neg \varphi \ 
 | \ \varphi_1 \wedge \varphi_2 \ 
 | \  {\bf P}_{\sim p} [\psi]\\
{\text {path formula }}  \psi &::=&
 {\bf X} \varphi \ 
 | \  \varphi_1 {\bf U}^{I} \varphi_2 \
 | \ {\bf Q}_{\bowtie q}^{I} \tuple{\varphi_1 | \varphi_2}
\end{eqnarray*}
against finite-state MCs.

The outline of the algorithm is similar to that for PCTL/CSL \cite{hansson1994, baier2003,kwiatkowska2007}.
We recursively compute a set $Sat(\varphi)$ of states satisfying $\varphi$
 from sets of states satisfying subformulae of $\varphi$.
\begin{eqnarray*}
Sat(a) &=& \{ s \in S | s \in L(a)\} \\
Sat(\neg \varphi) &=& S \setminus Sat(\varphi) \\
Sat(\varphi_1 \wedge \varphi_2) &=& Sat(\varphi_1) \cap Sat(\varphi_2)\\
Sat(P_{\sim p}[\psi]) &=& \{ s \in S | Prob^{\mathcal{D}/\mathcal{C}}(\psi)(s) \sim p \}
\end{eqnarray*}
where $Prob^{\mathcal{D}/\mathcal{C}}(\psi)$
 is the vector of occurrence probabilities of paths satisfying $\psi$ for each starting state in discrete-time/continuous-time.

In this paper,
 we indicate only how to compute $Prob^{\mathcal{D}/\mathcal{C}}(\psi)$ for 
 the case $\psi={\bf Q}_{\bowtie q}^{I} \tuple {\varphi_1 |\varphi_2}$.
For $\psi={\bf X} \varphi$ or $\psi=\varphi_1 {\bf U}^{I} \varphi_2$,
 we can use procedure for PCTL/CSL.
We assume that $Sat(\varphi_1)$ and $Sat(\varphi_2)$ are already computed,
 and
 that an interval $I$ is either of the form $[k,k']$ ($k' \neq \infty$) or $[k,\infty)$,
 because all intervals of $\mathbb{N}$ can be represented in one of these forms,
 and $Prob^{\mathcal{C}}({\bf Q}_{\bowtie q}^{I} \tuple {\varphi_1 |\varphi_2 })$ is equal to
 $Prob^{\mathcal{C}}({\bf Q}_{\bowtie q}^{[\inf I,\sup I]} \tuple {\varphi_1 |\varphi_2 })$ for $I$ such that $\inf I \neq \sup I$.

\paragraph{${\bf P}_{\sim p} [{\bf Q}_{\bowtie q}^{I} \tuple{ \varphi_1 | \varphi_2 }]$ for DTMCs.}
If $I=[k,k']$ ($k' \in \mathbb{N}$),
 we compute the occurrence probability of a path
 by counting the number of states satisfying $\varphi_2$ and $\varphi_1\wedge\varphi_2$
 in the interval $[k,k']$ on the path.
Let the vector $v_{j,i}^h(s)$ be the occurrence probability of a path
 starting from $s$,
 visiting states in $Sat(\varphi_2)$ $i$ times 
 and
 states in $Sat(\varphi_1) \cap Sat(\varphi_2)$ $j$ times,
 within $h$ steps:
\begin{eqnarray*}
v_{j,i}^0(s)&=&
\begin{cases}
1 & \text{if } (i=0, j=0 \text{ and } s \not\in Sat(\varphi_2)) \text{ or}\\
 & \quad (i=1, j=0 \text{ and } s \in Sat(\varphi_2) \setminus Sat(\varphi_1)) \text{ or}\\
 & \quad (i=1, j=1 \text{ and } s \in Sat(\varphi_1) \cap Sat(\varphi_2)),\\
0 & \text{otherwise.}
\end{cases}\\
v_{j,i}^h(s)&=&
\begin{cases}
P(s,-) \cdot v_{j-1,i-1}^{h-1} & \text{if }s \in Sat(\varphi_1) \cap Sat(\varphi_2), \\
P(s,-) \cdot v_{j,i-1}^{h-1} & \text{if } s \in Sat(\varphi_2) \setminus Sat(\varphi_1), \\
P(s,-) \cdot v_{j,i}^{h-1} & \text{otherwise.}
\end{cases}
\end{eqnarray*}
where $P(n,-)$ is the $n$-th row vector of the transition probability matrix $P$.

Here
 $Prob^{\mathcal{D}}({\bf Q}_{\bowtie q}^{[k,k']} \tuple{ \varphi_1 | \varphi_2 })$ is the probability
 of satisfying ${\bf Q}_{\bowtie q}^{[0,k-k']} \tuple{ \varphi_1 | \varphi_2 }$ after $k$-steps,
 and
 the $k$-step transition probability matrix is computed by $P^k$.
Hence,
 we can compute $Prob^{\mathcal{D}}({\bf Q}_{\bowtie q}^{[k,k']} \tuple{ \varphi_1 | \varphi_2 })$ as follows:
\begin{eqnarray*}
Prob^{\mathcal{D}}({\bf Q}_{\bowtie q}^{[k,k']} \tuple{ \varphi_1 | \varphi_2 }) = 
P^{k}
\cdot
\sum_{i=0}^{k'-k+1}
\sum_{
\substack{j \leq i\\
 i>0 \Rightarrow j \bowtie i \cdot q}
}
 v_{j,i}^{k'-k}
\end{eqnarray*}

If $I = [k,\infty)$ (unbounded),
 the basic idea is similar to the algorithm for the ${\bf S}$ operator in CSL \cite{baier2003,kwiatkowska2007}.
Each path in a finite-state MC
 has to reach one bottom strongly connected component (BSCC) $B$
 ($B$ is a strongly connected component, and $s \in B$ cannot reach $s' \not\in B$).
BSCCs are computed by Tarjan's Algorithm ($\mathcal{O}(|S|)$).
For a non-BSCC $A$ and BSCCs $B_1, \ldots, B_n$,
 let the matrix $P$ be reordered as
\begin{eqnarray*}
\begin{bmatrix}
P_A		& P_{AB_1} 	& \ldots		& \ldots		& P_{AB_n} \\
{\bf 0}	& P_{B_1}		&  {\bf 0}		& \ldots		& {\bf 0} \\
\vdots	& \ddots		&  \ddots 		& \ddots		& \vdots \\
\vdots	& 			&  \ddots	 	& \ddots		& {\bf 0} \\
{\bf 0}	& \cdots		&  \cdots		& {\bf 0}		& P_{B_n} \\
\end{bmatrix}
\end{eqnarray*}
where $P_{XY}$ is a partial transition matrix from $X$ to $Y$ of the transition probability matrix $P$ with $X,Y \subseteq S$
 (we denote a partial transition matrix $P_{XX}$ by $P_{X}$).

For each path reaching the BSCC $B_i$,
 the occurrence frequency of states converges to the limit distribution $\pi_{B_i}$
 depending on $B_i$.
 $\pi_{B_i}$ can be computed as the unique solution of the system of linear equations:
\begin{eqnarray*}
\pi_{B_i} P_{B_i} = \pi_{B_i} \text{ and } \pi_{B_i} \vec{{\bf 1}}= 1
\end{eqnarray*}
where $\vec{{\bf 1}}$ is the vector in which all elements are $1$.

If the BSCC $B_i$ has $s \in Sat(\varphi_2)$,
 the global frequency of $\varphi_1$ under $\varphi_2$ converges according to the limit distribution $\pi_{B_i}$.
Otherwise,
 the global frequency is determined by the local frequency before reaching $B_i$.
Then we compute the probability vector $r_A, r_{B_1}, \ldots, r_{B_n}$
 of reaching BSCCs for which the global frequency of $\varphi_1$ under $\varphi_2$ obeys the bound $\bowtie q$.
For the BSCC $B_i$ having state $s \in Sat(\varphi_2)$,
\begin{eqnarray*}
r_{B_i}
= 
\begin{cases}
\vec{\bf 1} & \text{if }
B_i \cap Sat(\varphi_2)\neq \emptyset \text{ and }
\displaystyle\frac{\sum_{s \in B_i \cap Sat(\varphi_1) \cap Sat(\varphi_2)} \pi_{B_i}(s)}
{\sum_{s \in B_i \cap Sat(\varphi_2)} \pi_{B_i}(s)}
 \bowtie q,\\
\vec{\bf 0} & \text{otherwise.}
\end{cases}
\end{eqnarray*}
For the non-BSCC $A$, $r_{A}$ can then be computed as the unique solution of the system of linear equations:
\begin{eqnarray*}
(P_{A} - {\bf I}) r_A
= -
\sum_{0\leq i \leq n}
P_{A{B_i}} r_{B_i}.
\end{eqnarray*}
Finally,
 we compute the probability of reaching BSCCs having no state $s \in Sat(\varphi_2)$ and satisfying the bound $\bowtie q$.
In a manner similar to the procedure used for $v_{j,i}^h$,
 we compute the occurrence probability of a path
 by counting the number of states satisfying $\psi_2$ and $\psi_1\wedge\psi_2$
 until reaching BSCCs that have no state $s \in Sat(\varphi_2)$ on the path.
Let
 the vector $u_{j,i}^h(s)$ be the occurrence probability of a path
 starting from $s$,
 visiting states in $Sat(\varphi_2)$ $i$ times,
 states in $Sat(\varphi_1) \cap Sat(\varphi_2)$ $j$ times,
 and
 states in $\bigcup_{B_i \cap Sat(\varphi_2) = \emptyset} B_{i}$ in $h$ steps the first time,
 within $h$ steps.
\begin{eqnarray*}
u_{0,0}^0(s)=
\begin{cases}
1 & \text{if } s \in \displaystyle\bigcup_{B_i \cap Sat(\varphi_2) = \emptyset} B_{i},\\
0 & \text{otherwise}.
\end{cases},\quad
u_{j,i}^h(s)=
\begin{cases}
0 & \text{if } s \not\in A,\\
P(s,-) \cdot u_{j-1,i-1}^{h-1} & \text{if } s \in Sat(\varphi_1) \cap Sat(\varphi_2) \cap A,\\
P(s,-) \cdot u_{j,i-1}^{h-1} & \text{if } s \in (Sat(\varphi_2) \setminus Sat(\varphi_1)) \cap A,\\
P(s,-) \cdot u_{j,i}^{h-1} & \text{otherwise}.
\end{cases}
\end{eqnarray*}
The reason
 $u_{j,i}^h(s)=0$ if $s \not\in A$ for $h>0$
 is
 that $s \not\in A$ cannot reach BSCCs having no state $s \in Sat(\varphi_2)$ in $h$ steps the first time.

The probability of reaching BSCCs having no state $s \in Sat(\varphi_2)$ and satisfying the bound $\bowtie q$
 can be obtained analytically as the infinite sum of $u_{j,i}^h(s)$ for $h=0$ to $\infty$,
 because the number of steps required to reach BSCCs from states in the non-BSCC $A$ is unbounded.
However,
 we can adequately approximate the true probability
 for large $h$ (see Section \ref{complexity_numerical}).
Thus
 we can compute $Prob^{\mathcal{D}}({\bf Q}_{\bowtie q}^{[k,\infty)} \tuple{ \varphi_1 | \varphi_2 })$ as follows:
\begin{eqnarray*}
Prob^{\mathcal{D}}({\bf Q}_{\bowtie q}^{[k,\infty)} \tuple{ \varphi_1 | \varphi_2 }) = 
P^k \cdot (
\begin{bmatrix}
r_A^T,
r_{B_1}^T,
\ldots,
r_{B_n}^T
\end{bmatrix}^T
+
\sum_{h=0}^{\infty}
\sum_{i=0}^{h+1}
\sum_{\substack{j \leq i\\ i>0 \Rightarrow j \bowtie i \cdot q}} u_{j,i}^h
).
\end{eqnarray*}
where the superscript $^T$ means transposition of a vector.
\paragraph{${\bf P}_{\sim p} [ {\bf Q}_{\bowtie q}^{I} \tuple{\varphi_1|\varphi_2}]$  for CTMCs.}
On a uniformized DTMC $\text{unif}_{\lambda}(\mathcal{C}) = (S,\bar{s}, P, L)$ of the input CTMC $\mathcal{C}$,
 the occurrence probability of sequences $s_0 s_1 \ldots$ of states can be captured
 by the techniques for DTMCs.
Therefore,
 the remainder is the occurrence probability of sequences $t_0 t_1 \ldots$ of spent times
 such that
 the ratio of
 the total spent time in states
 obeys the bound $\bowtie q$
 on the path, for the uniformization rate $\lambda$.

Consider a simple case that
 $i$ states ($s_0, \ldots ,s_i$, $i-1$ transitions) are in $[0,k]$,
 the number of transitions is $l$ ($j \leq l < i$) in $[0,qk]$
 and
 $i-l-1$ in the rest of the interval $(qk, k]$ on the path.
In this case,
 the total of $t_0$ to $t_{j-1}$ is less than $q \cdot k$
 and
 the occurrence probability of a sequence of spent times $t_0 \ldots t_i$ is
\begin{eqnarray*}
\rho(l;\lambda q k) \cdot \rho(i-l-1;\lambda (1-q) k)
 &=&
 e^{-\lambda q k} \cdot \frac{(\lambda q k)^l}{l!}
 \cdot
 e^{-\lambda (1-q) k} \cdot \frac{(\lambda (1-q) k)^{i-l-1}}{(i-l-1)!} \\
 &=&
 \rho(i-1;\lambda k) \cdot \frac{(i-1)! \cdot q^l \cdot (1-q)^{i-l-1}}{l! \cdot (i-l-1)!}.
\end{eqnarray*}
As above,
 the occurrence probability of a sequence of spent times obeying the given frequency bound
 depends on only
 the numbers of states satisfying subformulae in the interval of interest,
 and it can be computed using
 the binomial distribution,
 because
 each spent time is independent and exponentially distributed with parameter $\lambda$,
 and the Poisson probability
 $\rho(i-1;\lambda k)$ is the occurrence probability of $i-1$ transitions in $[0,k]$.
Under the other conditions,
 we can obtain similar results.
%
Hence,
 the conditional probability $B_{\bowtie q}(j,i)$ of satisfying the frequency bound $\bowtie q$,
 when
 the numbers of states satisfying $\varphi_2$
 and
 $\varphi_1\wedge\varphi_2$ in the interval $I$ are $i$ and $j$ respectively,
 is given by:
\begin{eqnarray*}
B_{\bowtie q}(j,i) =
\begin{cases}
1 & \text{if } i=0 \text{ or } (i=j \text{ and } 1\bowtie q) \text{ or } (j=0 \text{ and } 0\bowtie q), \\
\displaystyle\sum_{l=j}^{i-1} \frac{(i-1) !  \cdot q^l \cdot (1-q)^{i-l-1}}{ l! \cdot (i-l-1)!} & \text{if } 0<j<i, 0<q<1 \text{ and } \bowtie \in \{<, \leq\},\\
\displaystyle\sum_{l=0}^{j-1} \frac{(i-1) !  \cdot q^l \cdot (1-q)^{i-l-1}}{ l! \cdot (i-l-1)!} & \text{if } 0<j<i, 0<q<1\text{ and } \bowtie \in \{>, \geq\},\\
0 & \text{otherwise.}
\end{cases}
\end{eqnarray*}
Here
 $Prob^{\mathcal{D}}({\bf Q}_{\bowtie q}^{[k,k']} \tuple{ \varphi_1 | \varphi_2 })$ is the probability
 of satisfying \allowbreak ${\bf Q}_{\bowtie q}^{[0,k-k']} \tuple{ \varphi_1 | \varphi_2 }$ after $k$ time units,
 analogous to the DTMC case,
 and
 the transient probability for $k$ time units is $\Pi_k^{\mathcal{C}}$.
Therefore,
 for a bounded interval $I = [k,k']$,
\begin{eqnarray*}
Prob^{\mathcal{C}}({\bf Q}_{\bowtie q}^{[k,k']} \tuple{ \varphi_1 | \varphi_2 }) = 
\begin{cases}
\Pi_k^{\mathcal{C}} \cdot(
v_{0,0}^0 \cdot
B_{\bowtie q}(0,0)
+
v_{0,1}^0 \cdot
B_{\bowtie q}(0,1)
+
v_{1,1}^0 \cdot
B_{\bowtie q}(1,1))
& \text{if } k=k', \\
\Pi_{k}^{\mathcal{C}} \cdot
\displaystyle\sum_{h=0}^{\infty} \rho( h ; \lambda \cdot (k'-k))
 \cdot
\sum_{i=0}^{ h+1}
\sum_{j=0}^{i}
v_{j,i}^h \cdot
B_{\bowtie q}(j,i)
& \text{otherwise.}
\end{cases}
\end{eqnarray*}
In the numerical computation,
 this infinite sum for $h=0$ to $\infty$ can also be truncated as the computation of the transient probability $\Pi_k^{\mathcal{C}}$.

For an unbounded interval $I=[k,\infty)$,
 we can apply a routine similar to that used for DTMCs.
The difference is that we must consider
 the cumulative binomial probability $B_{\bowtie q}(j,i)$ for $u_{j,i}^h$,
 and
 the transient probability $\Pi_k^{\mathcal{C}}$ for $k$ time units instead of $P^k$.
\begin{eqnarray*}
Prob^{\mathcal{C}}({\bf Q}_{< q}^{[k,\infty)} \tuple{ \varphi_1 | \varphi_2 }) = 
\Pi_k^{\mathcal{C}} \cdot (
\begin{bmatrix}
r_A^T,
r_{B_1}^T,
\ldots,
r_{B_n}^T
\end{bmatrix}^T
+
\sum_{h=0}^{\infty}
\sum_{i=0}^{h+1} 
\sum_{j=0}^{i}
u_{j,i}^h \cdot
B_{\bowtie q}(j,i)
).
\end{eqnarray*}
%
%
%
\subsection{Model checking the bounded LTL-like fragment of PFTL}\label{for_statistical_model}
In this section,
 we introduce a statistical model-checking algorithm for infinite-state MCs and the bounded LTL-like fragment of PFTL:
\begin{eqnarray*}
{\text {state formula }} \varphi &::=& {\bf P}_{\sim p} [\psi]\\
{\text {path formula }}  \psi &::=&
 a \ 
 | \ \neg \psi \ 
 | \ \psi_1 \wedge \psi_2 \ 
 | \ \psi_1 {\bf U}^{I} \psi_2 \
 | \ {\bf Q}_{\bowtie q}^{I} \tuple{\psi_1 | \psi_2}
\end{eqnarray*}
where $p \in (0,1)$ and $I$ is a bounded interval of $\mathbb{N}$ for discrete time
 (or of $\mathbb{R}_{\geq 0}$ for continuous time).

Because
 it is difficult to check exactly for a bounded LTL-like fragment formula in PFTL,
 we develop a statistics-based approximation model-checking algorithm.
This techniques will provide us with useful information in many cases,
 even if it is not a strict model-checking procedure. 
In this approach,
 we sample finite prefix sequences of the paths of an input MC by probabilistic simulation
 and
 statistically determine whether or not the input MC satisfies an input formula by using the sample.
We apply the sequential probability ratio test (SPRT) \cite{wald1945}
 to model checking,
 as was done in \cite{younes2002} for CSL.
%
%
\subsubsection{Sequential probability ratio test}
The SPRT is a sequential hypothesis test
 developed by Wald \cite{wald1945}.
In a sequential test,
 the sample size is not fixed:
 observations are sequentially generated
 until the sample data indicate
 which hypothesis to supported under predesigned conditions.
In SPRT,
 we preset
  the type I error rate $\alpha>0$,
  the type II error rate $\beta>0$, and
  the indifference region width $2 \delta>0$.
For a formula ${\bf P}_{\sim p} [\psi]$ ($p\pm\delta \in (0,1)$),
 we test the null hypothesis $H_0$: $\hat{p} > p+\delta$
 against the alternative hypothesis $H_1$: $\hat{p} < p-\delta$,
 where $\hat{p}$ is the true value of the occurrence probability of paths satisfying $\psi$.
If the hypothesis $\hat{p}=\theta$ is true,
 the number $m$ of paths satisfying $\psi$ for a sample size $n$ is
 binomially distributed ${n! \theta^m (1-\theta)^{n-m}}/{(m! (n-m)!)}$.
Conversely,
 this value represents the likelihood of the hypothesis $\hat{p}=\theta$
 if we observe that $m$ paths satisfy $\psi$ for a sample size $n$.
Therefore,
 the likelihood ratio $\Lambda$ of $H_0$ to $H_1$ for a sample $\{\omega_1,\ldots,\omega_n\}$ is:
\begin{eqnarray*}
\Lambda(\{\omega_1,\ldots,\omega_n\}) =  
\frac
{(p+\delta)^m (1-(p+\delta))^{n-m}}
{(p-\delta)^m (1-(p-\delta))^{n-m}}
\end{eqnarray*}
where $m= |\{ \omega_i | \omega_i \models \psi \}|$.

Here
 $H_0$ is more likely than $H_1$ for a given sample if the likelihood ratio is greater than $1$
 and
 $H_1$ is more likely than $H_0$ for the sample if the likelihood ratio is less than $1$.
For an observed sample $\{\omega_1, \ldots ,\omega_n\}$ and error rates $\alpha$ and $\beta$,
 the next action is determined as follows:
\begin{eqnarray*}
\begin{cases}
\text{Accept } H_0 & \text{if } \Lambda (\{\omega_1, \ldots ,\omega_n\}) > (1-\beta) / \alpha,\\
\text{Accept } H_1 & \text{if } \Lambda (\{\omega_1, \ldots ,\omega_n\}) < \beta/(1-\alpha),\\
\text{Observe and add } \omega_{n+1} \text{ to the sample} & \text{otherwise.}
\end{cases}
\end{eqnarray*}
As a result,
 the probability of accepting the hypothesis $H_0$ is
 at least $1-\alpha$ if $\hat{p} > p+\delta$,
 and
 at most $\beta$ if $\hat{p} < p-\delta$.
If $|\hat{p} - p | < \delta$,
 the hypotheses are indifferent at error rates $\alpha$ and $\beta$.

%
%
%
\subsubsection{Satisfaction checking for bounded path formulae against paths}
To carry out a test,
 we must check $\omega \models \psi$ 
 for a sample path $\omega$ and a bounded formula $\psi$ with the total boundary $k_{total}$.
Whether or not  $\omega \models \psi$ does not depend on
 the suffix after $k_{total}$ steps/time-units of $\omega$.
For the finite prefix on $[0,k_{total}]$ of $\omega$,
 we recursively compute an ordered set $SatInt_{\omega}(\psi)$ of subintervals satisfying $\psi$ in $[0,k_{total}]$,
 using ordered sets of subintervals satisfy subformulae of $\psi$.
We can then derive $\omega \models \psi$ if there exists $I \in SatInt_{\omega}(\psi)$ such that $0 \in I$.

We assume that $SatInt_{\omega}(\psi_1)$ and $SatInt_{\omega}(\psi_2)$
 are already computed and merged.
By writing
 $SatInt_{\omega}(\psi)$
 $ = \{I_i, \ldots, I_n\}$,
 we mean that
 the set $\{I_i, \ldots, I_n\}$ satisfies
 $I_i \cap I_{i+1} = \emptyset$,
 $\sup I_{i} \leq \inf I_{i+1}$ and
 $\sup I_{i} = \inf I_{i+1} \Rightarrow \sup I_i \not\in I_{i}, I_{i+1}$.
In this paper,
 we do not include an algorithm for DTMCs,
 because
 the structure of a discrete-time path is simple,
 and it is not worthwhile to pursue the matter.

\paragraph{$a \in AP$ for CTMCs.}
For an atomic proposition $a \in AP$,
 the set of intervals satisfying $a$ is determined immediately by the labeling function $L$.
Therefore,
$SatInt_{\omega}(a) = 
\{
[ \sum_{j=0}^{i-1} time(\omega,j), \sum_{j=0}^{i} time(\omega,j) )
| a \in L(\omega(i))
 \}$.
\paragraph{$\neg \psi_1$ for CTMCs.}
$SatInt_{\omega}(\neg\psi_1)$
 is a set of intervals complementary to the union of intervals in $SatInt_{\omega}(\psi_1)$ in $[0, k_{total}]$.
Therefore,
 for $SatInt_{\omega}(\psi_1) = \{ I_1, \ldots I_n\}$,
$SatInt_{\omega}(\neg \psi_1) =
\{ [0, \inf I_1] \setminus I_1,  [\sup I_n, k_{total}] \setminus I_n\}
 \cup \bigcup_{i=1}^{n-1} \{ ([\sup I_i, \inf I_{i+1}] \setminus I_i) \setminus I_{i+1} \}
 $.
\paragraph{$\psi_1 \wedge \psi_2$ for CTMCs.}
$SatInt_{\omega}(\psi_1 \wedge \psi_2)$ is
 a set of intervals intersecting each element of $SatInt_{\omega}(\psi_1)$ and each element of $SatInt_{\omega}(\psi_2)$.
Therefore,
 for $SatInt_{\omega}(\psi_1) = \{ I_1, \ldots I_n\}$ and $SatInt_{\omega}(\psi_2) = \{ J_1, \ldots J_m\}$,
$SatInt_{\omega}(\psi_1 \wedge \psi_2) = \bigcup_{i=1}^{n} \bigcup_{j=1}^{m} \{  I_i \cap J_j \}$.
\paragraph{$\psi_1 {\bf U}^{I} \psi_2$ for CTMCs.}
Let $SatInt_{\omega}(\psi_1) = \{ I_1, \ldots I_n\}$ and $SatInt_{\omega}(\psi_2) = \{ J_1, \ldots J_m\}$.
For time points $t \in I_i \cup \{\inf I_i\}$ and $t' \in J_j$ such that $t<t'$,
 there exists $I' \in SatInt_{\omega}(\psi_1 {\bold U}^I \psi_2)$ such that $t \in I'$ if $(t, t') \subseteq I_i$ and $t' \in I + t$.
In this case, 
 $t'$ is in $(I_i \cup \sup I_i ) \cap J_j$ ($=Y_{i,j}$) and
 $t$ is in $X_{i,j}$ where
 $\inf X_{i,j} = \inf Y_{i,j} - \sup I$,
 $\sup X_{i,j} = \sup Y_{i,j} - \inf I$,
 $(\inf Y_{i,j} \in Y_{i,j} \wedge \sup I \in I )\Leftrightarrow \inf X_{i,j} \in X_{i,j}$ and
 $(\sup Y_{i,j} \in Y_{i,j} \wedge \inf I \in I)\Leftrightarrow \sup X_{i,j} \in X_{i,j}$.
In addition,
 $SatInt_{\omega}(\psi_2) \subseteq SatInt_{\omega}(\psi_1 {\bold U}^I \psi_2)$ if $0 \in I$.
Therefore,
\begin{eqnarray*}
SatInt_{\omega}(\psi_1 {\bold U}^I \psi_2) =
\bigcup_{j=1}^{m} 
\bigcup_{i=1}^{n} 
\{  X_{i,j} \cap (I_i \cup \{\inf I_i\})\}
 \cup
\begin{cases}
SatInt_{\omega} (\psi_2) & \text{if } 0 \in I,\\
\emptyset & \text{otherwise.}
\end{cases}
\end{eqnarray*}
\paragraph{${\bf Q}_{\bowtie q}^{I} \tuple{\psi_1 | \psi_2} $ for CTMCs.}
If $I=[0,0]$,
 ${\bold Q}_{\bowtie q}^{[0,0]} \tuple{\psi_1 | \psi_2}$ is just a conditional statement.
Therefore,
$\allowbreak SatInt_{\omega}({\bold Q}_{\bowtie q}^{[0,0]} \tuple{\psi_1 | \psi_2})$ is equal to
$SatInt_{\omega}(\psi_2 \rightarrow \psi_1)$ if $1\bowtie q$,
$SatInt_{\omega}(\psi_2 \rightarrow \neg\psi_1)$ otherwise.

If $\inf I >0$,
$\mathcal{C}, \omega \models {\bf Q}_{\bowtie q}^{I} \tuple{\psi_1 | \psi_2} $
$\Leftrightarrow$
$\mathcal{C}, \omega^{\inf I} \models {\bf Q}_{\bowtie q}^{I - \inf I} \tuple{\psi_1 | \psi_2} $ by Definition \ref{sem_for_ctmc}.
Therefore,
 if $\inf I >0$,
$SatInt_{\omega}({\bf Q}_{\bowtie q}^{I} \tuple{\psi_1 | \psi_2})$
$=$
$\{ J - \inf I | J \in SatInt_{\omega}({\bf Q}_{\bowtie q}^{I - \inf I} \tuple{\psi_1 | \psi_2})$.

If $\inf I =0$ and $\sup I =k > 0$,
 $SatInt_{\omega}({\bold Q}_{\bowtie q}^{I} \tuple{\psi_1 | \psi_2})$ satisfy the property 
$J \in SatInt_{\omega}({\bold Q}_{\bowtie q}^{I} \tuple{\psi_1 | \psi_2})$
$\Leftrightarrow$
$J \in \{ t | f^I_{\tuple{\psi_1|\psi_2}} (t) \bowtie q\}$
 for any interval $J$.
Therefore,
 we determine $SatInt_{\omega}({\bf Q}_{\bowtie q}^{I - \inf I} \tuple{\psi_1 | \psi_2})$ 
 by analyzing $f^I_{\tuple{\psi_1|\psi_2}} (t)$.
First,
 for $SatInt_{\omega}(\psi_1 \wedge \psi_2) = \{ I_1, \ldots I_n\}$ and
 $SatInt_{\omega}(\psi_2) = \{ J_1, \ldots J_m\}$,
 we compute a set nondif$_{\omega}(\psi_1 | \psi_2)$ of candidates for non-differentiable
 points of $f^I_{\tuple{\psi_1|\psi_2}} (t)$.
\begin{eqnarray*}
\text{nondif}_{\omega}(\psi_1 | \psi_2) = 
\{0, k_{total}-k\} 
 \cup
\{ \inf I' -k,\inf I', \sup I'-k,\sup I' 
 | 
 I'  \in \{I_1, \dots, I_n, J_1, \dots, J_m\}
\}.
\end{eqnarray*}
Let $\{ t_1, \ldots, t_l \}$ be the ordered elements of nondif$_{\omega}(\tuple{\psi_1 | \psi_2})$.
The truth values of $\psi_2$ and $\psi_1\wedge\psi_2$ are unchanged in each interval $(t_i, t_{i+1})$ and $(t_i, t_{i+1})+k$,
 because
 if their truth values did change,
 there would have to be other non-differentiable points between $t_i$ and $t_{i+1}$.
Thus
 $f^I_{\tuple{\psi_1|\psi_2}}(t)$ is
 monotonically increasing,
 monotonically decreasing,
 fixed, or
 undefined
 in the interval $(t_i, t_{i+1})$.
In addition,
 $f^I_{\tuple{\psi_1|\psi_2}}(t)$ is equal to $f^{(\inf I, \sup I)}_{\tuple{\psi_1|\psi_2}}(t)$ for $t \in (t_i,t_{i+1})$.

Hence,
 for a non-differentiable time point $t_i$,
 $[t_i,t_i] $
 $\in$
 $SatInt_{\omega}({\bf Q}_{\bowtie q}^{I} \tuple{\psi_1 | \psi_2})$
 if $f^I_{\tuple{\psi_1|\psi_2}} (t_i) \bowtie q$.
For an interval $(t_{i}, t_{i+1})$ between non-differentiable time points,
 we determine whether or not
 $(t_{i}, t_{i+1})$, or a subinterval of it, is in $SatInt_{\omega}({\bf Q}_{\bowtie q}^{I} \tuple{\psi_1 | \psi_2})$
 as follows.
For $\mathcal{L}^{J}_{\psi}(t) = \mathcal{L}(\bigcup_{I' \in SatInt(\psi)} I' \cap (J+t))$:
\begin{enumerate}
\item[1.] If
$\mathcal{L}^{(0,k)}_{\psi_2}(t_i)=0$ and $\mathcal{L}^{(0,k)}_{\psi_2}(t_{i+1})=0$,
$\psi_2$ and $\psi_1\wedge\psi_2$ do not hold on an interval $\subseteq (t_i,t_{i+1})$ with positive time length.
Therefore,
 if $f^{(0,k]}_{\tuple{\psi_1|\psi_2}}(t_i)$ ($=f^{(0,k)}_{\tuple{\psi_1|\psi_2}}(t')$ for $t_i<t'<t_{i+1}$)
 is undefined or obeys the bound $\bowtie q$,
 then
 $(t_i,t_{i+1}) \in SatInt_{\omega}({\bold Q}_{\bowtie q}^{I} \tuple{\psi_1 | \psi_2} )$.
\item[2.] If
 $\mathcal{L}^{(0,k)}_{\psi_2}(t_i)=0$ and $\mathcal{L}^{(0,k)}_{\psi_2}(t_{i+1})>0$,
 $f^I_{\tuple{\psi_1|\psi_2}}(t)$ is fixed and equal to $f^{(0,k)}_{\tuple{\psi_1|\psi_2}}(t_{i+1})$ in the interval $(t_i,t_{i+1})$.
Therefore,
 if $f^{(0,k)}_{\tuple{\psi_1|\psi_2}}(t_{i+1})$ obeys the bound $\bowtie q$,
 then
 $(t_i,t_{i+1}) \in SatInt_{\omega}({\bold Q}_{\bowtie q}^{I} \tuple{\psi_1 | \psi_2})$.
\item[3.] If
 $\mathcal{L}^{(0,k)}_{\psi_2}(t_i)>0$ and $\mathcal{L}^{(0,k)}_{\psi_2}(t_{i+1})=0$,
 $f^I_{\tuple{\psi_1|\psi_2}}(t)$ is fixed and equal to $f^{(0,k)}_{\tuple{\psi_1|\psi_2}}(t_{i})$ in the interval $(t_i,t_{i+1})$.
Therefore,
 if $f^{(0,k)}_{\tuple{\psi_1|\psi_2}}(t_{i})$ obeys the bound $\bowtie q$,
 then
 $(t_i,t_{i+1}) \in SatInt_{\omega}({\bold Q}_{\bowtie q}^{I} \tuple{\psi_1 | \psi_2})$.
\item[4.] If
 $\mathcal{L}^{(0,k)}_{\psi_2}(t_i)>0$ and $\mathcal{L}^{(0,k)}_{\psi_2}(t_{i+1})>0$,
 $f^{(0,k)}_{\tuple{\psi_1|\psi_2}}(t)$ is 
 monotonically increasing,
 monotonically decreasing, or
 fixed 
 in $(t_i,t_{i+1})$.
Therefore,
 if both $f^{(0,k)}_{\tuple{\psi_1|\psi_2}}(t_i)$ and $f^{(0,k)}_{\tuple{\psi_1|\psi_2}}(t_{i+1})$ obey the bound $\bowtie q$,
 then
 $(t_i,t_{i+1}) \in SatInt_{\omega}({\bold Q}_{\bowtie q}^{I} \tuple{\psi_1 | \psi_2})$.
Moreover,
 if either $f^{(0,k)}_{\tuple{\psi_1|\psi_2}}(t_i)$ or $f^{(0,k)}_{\tuple{\psi_1|\psi_2}}(t_{i+1})$ obeys the bound $\bowtie q$,
 then
 $(t_i, t_i')$ or $(t_i', t_{i+1}) \in SatInt_{\omega}({\bold Q}_{\bowtie q}^{I} \tuple{\psi_1 | \psi_2})$
 where $t_i'$ satisfies:
\begin{eqnarray*}
q = \displaystyle\frac{ \mathcal{L}^{(0,k)}_{\psi_1\wedge\psi_2}(t_i) + a(t'_i -t_i) }{\mathcal{L}^{(0,k)}_{\psi_2}(t_i) + b(t'_i -t_i) }
\end{eqnarray*}
with
$a=(\mathcal{L}^{(0,k)}_{\psi_1\wedge\psi_2}(t_{i+1})-\mathcal{L}^{(0,k)}_{\psi_1\wedge\psi_2}(t_i))$
$/(t_{i+1}-t_i)$
 and
$b=(\mathcal{L}^{(0,k)}_{\psi_2}(t_{i+1})-\mathcal{L}^{(0,k)}_{\psi_2}(t_i))$
$/(t_{i+1}-t_i)$.
\begin{eqnarray*}
t'_i =
t_i + 
\displaystyle\frac{q \cdot \mathcal{L}^{(0,k)}_{\psi_2}(t_i) - \mathcal{L}^{(0,k)}_{\psi_1\wedge\psi_2}(t_i)}
{a - q b}.
\end{eqnarray*}
In addition,
 because $f^I(t'_i)=q$,
$[t_i',t_i'] \in SatInt_{\omega}({\bold Q}_{\bowtie q}^{I} \tuple{\psi_1 | \psi_2})$ if $\bowtie \in \{\leq,\geq\}$.
\end{enumerate}
%
%
%
\subsection{Complexity}\label{complexity}
\subsubsection{Complexity of model checking for the CTL-like fragment of PFTL}\label{complexity_numerical}
For
a DTMC $\mathcal{D}=(S,\bar{s},P, L)$
 or a CTMC $\mathcal{C}=(S,\bar{s}, Q, L)$ (and its uniformized DTMC $\text{unif}_{\lambda}(\mathcal{C}) = (S,\bar{s},P, L)$)
 and a CTL-like fragment formula $\varphi$,
 the time complexity of model checking is
 linear in $|\varphi|$, which is the number of operators in $\varphi$,
 and
 polynomial in $|S|$, which is the complexity of the recursive procedure for each operators.
Except for a ${\bf Q}$ formula,
 the time complexity is the same to that for PCTL/CSL \cite{hansson1994, baier2003}.
For each bounded ${\bf P}_{\sim p}[{\bf Q}_{\bowtie q}^{I} \tuple{\varphi_1 | \varphi_2}]$ ($\sup I < \infty$),
 computing the sum of vectors $v_{j,i}^h$ takes $\mathcal{O}( |S|^2 \cdot k^3)$ time,
 where $k=\sup I$ for DTMCs or $k = \lambda \cdot \sup I$ for CTMCs.
For each unbounded ${\bf P}_{\sim p}[{\bf Q}_{\bowtie q}^{I}  \tuple{\varphi_1 | \varphi_2}]$,
 computing the limit distributions $\pi_{B_1}, \ldots,\pi_{B_n}$ and
 the reachability vectors $r_A, r_{B_1}, \ldots, r_{B_n}$
 takes $\mathcal{O}(|S|^3)$ time,
 where $A$ and ${B_1}, \ldots,{B_n}$ are a non-BSCC and BSCCs of $P$, respectively.
If the input MC is reducible,
 an additional
 computation of the transient probability $\Pi_k^{\mathcal{C}}$ and the sum of vectors $u_{j,i}^h$
 takes $\mathcal{O}(|S|^2 \cdot k' + |S|^2 \cdot |\log e |^{-3})$ time,
 where
 $k'=\inf I$ for DTMCs or $k' = \lambda \cdot \inf I$ for CTMCs,
 and
 $e$ is the maximum absolute value
 of the eigenvalues of the partial matrix $P_A$ consisting the non-BSCC $A$ of $P$.
This is because
 the probability vector of reaching BSCCs within $\mathcal{O}(|\log e|^{-1})$-steps
 is sufficiently close to the probability vector of reaching BSCCs within an unbound number of steps.

\subsubsection{Complexity of model checking for the LTL-like fragment of PFTL}

The complexity of model checking for the LTL-like fragment of PFTL
 is divided into two parts,
 the complexity of the sample used in the testing
 and
 the complexity of the observations and satisfaction checking for a sample trace of path.
Regarding the sample size,
 approximations for the expected sample size are provided in \cite{wald1945, younes2006}.
This size depends on the chosen significance level
 and the difference between the query value $p$ and the true probability $\hat{p}_{\psi}$
 for an input formula ${\bf P}_{\sim p}[\psi]$.
However,
 this is not specific to our method, and the details of the expected size are omitted from this paper.
The observation of a sample path is just a probabilistic simulation,
 and its time complexity is 
 $\mathcal{O}(k_{total} \cdot \log |E| )$/$\mathcal{O}(\lambda \cdot k_{total}  \cdot \log |E|)$
 where
 $k_{total}$ is the total boundary of the input formula,
 $|E|$ is the number of transition choices of the input MC and
 $\lambda$ is the average exit rate of the input CTMC,
 for the input DTMC/CTMC.
For an input formula ${\bf P}_{\sim p}[\psi]$ and a DTMC,
 we need only count states satisfying subformulae for each operator.
Therefore,
 the satisfaction checking takes 
 $\mathcal{O}(k_{total} \cdot |\psi| )$ time,
 where $|\psi|$ is the size of $\psi$.
However,
 on a CTMC,
 the size of the set of intervals satisfying formulae
 is at worst twice that of the set of intervals satisfying subformulae,
 for each ${\bf Q}$ operator.
Thus,
 the satisfaction checking takes $\mathcal{O}(\lambda \cdot k_{total} \cdot |\psi| \cdot 2^{|\psi|_{{\bf Q}}} )$ time,
 where
 $|\psi|_{{\bf Q}}$ is the number of ${\bf Q}$ operators in $\psi$.
In practice,
 many intervals satisfying formulae are merged,
 because
 each spent time on a state is exponentially distributed with the exit rate of the state
 and
 the probability of generating a bad sample path by probabilistic simulation is negligible. 
%
%
%
\section{Conclusions and future directions}\label{last}
We introduced the frequency operator ${\bf Q}$ and
 defined the syntax and semantics of PFTL.
PFTL has rich expressiveness, and it is difficult to develop a model checker for the full logic.
However,
 we developed
 a numerical model-checking algorithm
 for the CTL-like fragment of PFTL against finite-state MCs,
 and
 a statistical model-checking algorithm
 for the bounded LTL-like fragment of PFTL against infinite-state MCs.
The statistical model-checking is not strict,
 but
 we anticipate that it will provide useful information in many cases.
Especially,
 it is worth noting that
 the ${\bf Q}$ operator can, in a sense, express
 a conditional probability between path formulae,
 without path quantifications.
%

Our extension is based on an intuitive idea for describing a property of a behavior, especially in a probabilistic system.
Although,
 it is difficult to strictly check a model for the logic, and also the non-probabilistic version of PFTL,
 because it is intractable from the viewpoint of automata theory.
Therefore,
 it will be necessary to find treatable and useful fragments of the logic and classes of restricted models.
This is one future direction of our research.
Another future direction is to provide approximate model-checking against more complex systems,
 or for further extended logics.
In this paper,
 we have assumed that our model is an MC.
Nevertheless,
 we can apply
 this type of approximate model-checking via statistical methods
 to more general stochastic processes,
 e.g., systems of stochastic ordinary differential equations (continuous states and continuous transitions),
 because we can directly use discretized traces of paths obtained from stochastic simulations.
Also,
 it is not difficult to check whether or not a sample path satisfies a bounded property
 such as ``$\varphi_2$ holds in the interval $[0,10]$ and $\varphi_1$ holds to more than 90\% of the time points until that point''
 (frequently $\varphi_1$ until $\varphi_2$).
%
%
\bibliographystyle{eptcs}
\bibliography{infinity2011}
%

%
\end{document}